# Does P-type Ohmic Contact Exist in $WSe_2$-metal Interfaces?

Yangyang Wang,[1,3,†] Ruoxi Yang,[1,4,†] Ruge Quhe,[1,5,6,†] Hongxia Zhong,[1,7,†] Linxiao Cong,[1] Meng Ye,[1] Zeyuan Ni,[1] Zhigang Song,[1] Jinbo Yang,[1,2] Junjie Shi,[1] Ju Li,[3] and Jing Lu[1,2,*]

[1] State Key Laboratory for Mesoscopic Physics and Department of Physics, Peking University, Beijing 100871, P. R. China

[2] Collaborative Innovation Center of Quantum Matter, Beijing 100871, P. R. China

[3] Department of Nuclear Science and Engineering and Department of Materials Science and Engineering, Massachusetts Institute of Technology, Cambridge, Massachusetts 02139, USA

[4] Department of Chemistry, University of Bath, Claverton Down, Bath BA2 7AY, UK

[5] State Key Laboratory of Information Photonics and Optical Communications, Beijing University of Posts and Telecommunications & School of Science, Beijing 100876, China

[6] Academy for Advanced Interdisciplinary Studies, Peking University, Beijing 100871, China

[7] Department of Physics, Washington University in St. Louis, St. Louis, Missouri 63130, USA

[†]These authors contributed equally to this work.

[*]Address correspondence to: jinglu@pku.edu.cn

## Abstract

Formation of low-resistance metal contacts is the biggest challenge that masks the intrinsic exceptional electronic properties of 2D $WSe_2$ devices. We present the first comparative study of the interfacial properties between ML/BL $WSe_2$ and Sc, Al, Ag, Au, Pd, and Pt contacts by using *ab initio* energy band calculations with inclusion of the spin-orbital coupling (SOC) effects and quantum transport simulations. The interlayer coupling tends to reduce both the electron and hole Schottky barrier heights (SBHs) and alters the polarity for $WSe_2$-Au contact, while the SOC chiefly reduces the hole SBH. In the absence of the SOC, Pd contact has the smallest hole SBH with a value no less than 0.22 eV. Dramatically, Pt contact surpasses Pd contact and becomes *p*-type Ohmic or quasi-Ohmic contact with inclusion of the SOC. Our study provides a theoretical foundation for the selection of favorable metal electrodes in ML/BL $WSe_2$ devices.

**Introduction**

Two dimensional (2D) transition-metal dichalcogenides (TMDs) are attracting much recent attention because they have a wide range of application prospects in electronics,[1-6] photoelectronics,[1, 7-9] spintronics,[10-12] and valleytronics.[11, 13-18] Among 2D TMDs, monolayer (ML) and bilayer (BL) MoS$_2$ and WSe$_2$ are probably most intensively studied. 2D WSe$_2$ distinguishes it from 2D MoS$_2$ mainly in two aspects: (1) 2D WSe$_2$ has a much enhanced spin-orbit coupling (SOC) due to heavier W and Se atoms.[11, 12] For example, the splitting of the valence band maximum (VBM) is about 0.15 eV in ML MoS$_2$ but is enhanced to 0.46 eV in ML WSe$_2$.[11] Therefore, 2D WSe$_2$ is more suitable for spintronics purpose. (2) 2D MoS$_2$ favors *n*-type doping,[2, 3, 19] whereas 2D WSe$_2$ prefers *p*-type doping as a result of much higher positions of the conduction band minimum (CBM) and the VBM.[5, 6] A *p-n* junction can be fabricated with 2D MoS$_2$ as *n* region and 2D WSe$_2$ as *p* region, and such a device has been reported recently, with excellent rectification behavior and rapid photoresponse.[16, 20]

A device often needs a contact with metal electrodes, and the formation of low-resistance metal contacts is the biggest challenge that masks the intrinsic exceptional electronic properties of 2D TMDs.[19] In the absence of a controllable and sustainable substitutional doping scheme, one has to rely on the work function of contact metals to inject appropriate types of carriers into the respective bands of 2D TMDs. Such metal-semiconductor contacts are often associated with a formation of finite Schottky barrier, which decreases the carrier injection efficiency. Apparently, decreasing Schottky barrier height (SBH) is critical to reach a high performance of a device, and a low resistance Ohmic contact with vanishing SBH is highly desirable. Unfortunately, the SBH does not simply depend on the difference between the intrinsic Fermi level ($E_f$) of a metal and the CBM or VBM of the semiconductor due to the complex Fermi level pinning, which renders the appearance of Ohmic contact rather difficult.[21, 22] There have been a lot of works to study 2D MoS$_2$-metal interfaces.[19, 23-27] However, no *n*-type low-resistance Ohmic contact has been revealed either experimentally or theoretically in 2D MoS$_2$ even with low work function metals such as Ti and Sc in terms of the recent reports [3, 19, 21, 28]

Compared with those about MoS$_2$, the studies of 2D WSe$_2$-metal interfaces are much limited. In the experimental aspect, Javey *et al.*'s measurement showed that high work function Pd forms the lowest resistance to the valence band of ML WSe$_2$ for hole transport, while lower function Ag, Ni, Au, Ti, and Gd have high SBH to both the valence band and the conduction band of ML WSe$_2$.[5] Banerjee *et al.*'s measurement showed that Al forms a *n*-type Schottky contact with ML WSe$_2$, but Ti, In, and Ag form *n*-type Ohmic contacts.[4] In the theoretical aspect, Schottky barrier is always present in ML WSe$_2$ and In, Ti, Al, Au, and Pd interfaces, where the SOC is not considered in the energy band calculations.[4, 19] Three fundamental issues arise: (1) Which elemental metal has the smallest hole SBH when contacted with WSe$_2$? (2) Whether *p*-type Ohmic contact is present in ML WSe$_2$-metal contacts? (3) What are the effects of the SOC on the SBH of ML WSe$_2$-metal contacts? The SBH depends on the difference between $E_f$ of the metal electrode and the band edge of the channel semiconductor in a device environment. Given a rise of 0.23 eV of the VBM due to the SOC splitting in ML WSe$_2$,[11] the ignorance of it in determining the hole SBH appears rather unacceptable unless existence of a full Fermi level pining during the SOC process.

Due to the strong interlayer coupling, BL WSe$_2$ has a reduced band gap compared with ML one (1.44 eV vs 1.67 eV at the density functional theory (DFT) level).[29] Therefore, a reduced SBH and thus a higher carrier injection efficiency can be expected in BL WSe$_2$-metal contacts, suggesting a better device performance of BL WSe$_2$ as the channel compared with ML WSe$_2$ given an identical gate voltage controllability. However, to the best of our knowledge, the interfacial properties of BL WSe$_2$-metal contacts have not been investigated.

In this Article, we provide a comparative study of the interfacial properties of ML and BL WSe$_2$ on several commonly used metals (Sc, Al, Ag, Au, Pd, and Pt), for the first time by using the DFT energy band calculation with inclusion of the SOC effects. We find that the interlayer coupling decreases both the electron and hole SBHs and even alters the polarity of WSe$_2$-Au contact. No Ohmic contact is revealed in the absence of the SOC, and Pd contact has the minimum hole SBH (0.22/0.27 eV for ML/BL WSe$_2$ case). However, when the SOC is included, ML and BL WSe$_2$-Pt interfaces dramatically have the minimum hole SBHs and actually become *p*-type Ohmic or quasi-Ohmic contact. A more reliable approach to treat the SBH is *ab initio* quantum transport simulation based on a two-probe model, which is also performed and gives a hole SBH similar to that of the energy band calculation for ML WSe$_2$-Pt interface in the absence of the SOC.

**Results and Discussion**

**Interface modeling and stability**

We use six layers of metal atoms (Sc in (0001) orientation and Al, Ag, Au, Pd and Pt in (111) orientation) to model the metal surface and adjust them to match the optimized lattice constant of ML and BL WSe$_2$. The calculated in-plane lattice constant of WSe$_2$ is $a = 3.29$ Å, in good agreement with the experimental value.[30] The (1 × 1) unit cell of Sc (0001) face is adjusted to the (1 × 1) unit cell of WSe$_2$, and (2 × 2) unit cells of Al, Ag, Au, Pd and Pt (111) faces are adjusted to the ($\sqrt{3} \times \sqrt{3}$) R30° unit cell of WSe$_2$ with mismatches within 4%. The in-plane lattice constant of the supercell is fixed during the relaxation. A vacuum buffer space of at least 20 Å is set to avoid spurious interaction between periodic images. The most stable ML WSe$_2$-metal contact geometries are obtained by optimizing the structures from different initial configurations. The initial configurations of BL WSe$_2$-metal interfaces are constructed in terms of AA′ stacking in WSe$_2$ (a $D_{3d}$ point group symmetry)[31] on the basis of the most stable ML WSe$_2$-metal interface configurations.

The most stable configurations of the ML WSe$_2$-metal interfaces are shown in Figure 1. The relative positions between ML WSe$_2$ and metal substrates along the interface directions are different for different metals. On Sc(0001) surface, the W atoms in the primitive cell sit above the top metal atoms, and the Se atoms sit above the second layer metal atoms (Figure 1c); On Al and Pt (111) surfaces, the W atoms in the supercell are all above the centers of the triangles formed by the fcc, hcp, and top sites, and the three pairs of Se atoms sit above the fcc, hcp, and top sites, respectively (Figure 1d); On Pd (111), the three W atoms in the supercell sit above the fcc, hcp, and top sites, respectively, and the Se atoms are all above the centers of the triangles formed by the fcc, hcp, and top sites (Figure 1e); On Ag and Au(111), the W and Se atoms are all above the centers of the triangles formed by the fcc, hcp, and top sites (Figure 1f). The most stable configurations of the BL WSe$_2$-metal interfaces are similar with the corresponding ML ones. The equilibrium interfacial distances $d_z$ in ML and BL WSe$_2$-metal contacts are insensitive to the WSe$_2$ layer number, varying from 2.271-2.959 Å (Table 1). The binding energy per interfacial W atom is defined as

$$E_b = (E_{WSe2} + E_{metal} - E_{WSe2\text{-}metal})/N_W \quad (1)$$

Where $E_{WSe2}$, $E_{metal}$, and $E_{WSe2\text{-}metal}$ are the relaxed energies for WSe$_2$, metal surface, and the combined system per supercell, respectively, and $N_W$ is the number of interface W atoms per supercell. $E_b$ ranges from 0.160 to 1.049 eV as listed in Table 1. Similar to the cases of MoS$_2$,[19] the adsorption of ML and BL WSe$_2$-metal surfaces can be classified into weak bonding (Al, Ag and Au contacts) with $E_b$ = 0.160-0.367 eV, medium bonding (Pt and Pd) with $E_b$ = 0.525-0.706 eV, and strong bonding (Sc) with $E_b$ = 0.918 (ML) and 1.049 (BL) eV according to the binding strength.

It is important to note that there are two possible interfaces to form Schottky barrier in a transistor as shown in Figure 1g: one is the source/drain interface (B) between the contacted WSe$_2$ and the metal surface in the vertical direction if the interaction between WSe$_2$ and the metal surface is weak, and the other is source/drain-channel (D) interface between the contacted WSe$_2$ and channel WSe$_2$ in the lateral direction if the interaction between WSe$_2$ and the metal surface is strong. Such a dual interface model has been employed in the recent MoS$_2$-, graphdiyne-, and ML phosphorene-metal contact studies.[19, 21, 22, 32] Compared with a single vertical interface model, which predicts an Ohmic contact between ML MoS$_2$ and Ti due to the strong band hybridization,[26] the dual interface model predicts a Schottky contact with an electron SBH of 0.22-0.35 eV.[19, 21] A significant electron SBH of 0.30-0.35 eV between ML MoS$_2$ and Ti is found experimentally,[19] justifying the dual interface model. Actually, the calculated (0.096 eV) and observed (0.065 eV) electron SBH for BL MoS$_2$-Ti[21, 33] based on the dual interface model also show a good consistency.

**Vertical SBH and overlap states**

The band structures of ML WSe$_2$ and the combined systems are shown in Figure 2. The ML WSe$_2$ has a band gap of 1.60 eV when the SOC is considered, consistent with the reported DFT value of 1.67 eV.[29] Both the valence and conduction bands of ML WSe$_2$ are strongly destroyed when contacted with Sc, resulting in an absent vertical Schottky barrier for ML WSe$_2$-Sc contact. The majority of the ML WSe$_2$ bands are still identifiable when contacted with Al, Ag and Au surfaces because the weak interaction. When contacted with Pt and Pd surfaces, the valence bands of ML WSe$_2$ are hybridized slightly with the $d$-bands of Pt and Pd, while the conduction bands are preserved well. Vertical Schottky barrier $\Phi_V^{e/h}$ for these weak or medium bonding cases (Figure 1g) can be obtained from the energy difference between $E_f$ of the interfacial system and the CBM (electron SBH) or VBM (hole SBH) of the contacted WSe$_2$. In ML WSe$_2$-Al, -Ag and -Au contacts, as shown in Figure 2, the vertical Schottky barriers are $n$-type with electron SBH of $\Phi_V^e$ = 0.62, 0.40, and 0.53 eV, respectively. While in ML WSe$_2$-Pd and -Pt contacts, the vertical Schottky barriers are $p$-type with hole SBH of $\Phi_V^h$ = 0.22 and 0.30 eV, respectively. The vertical electron/hole SBH in ML WSe$_2$-Au/-Pd contact (0.53/0.22 eV) is comparable with the one (0.70/0.35 eV) calculated by Banerjee et al.[19] in the absence of the SOC effects. The nearly midgap SBH character of Al contact ($\Phi_V^e$ = 0.62 eV and $\Phi_V^h$ = 0.79 eV ) is also consistent with the PDOS calculations of Banerjee et al..[4]

The band hybridization degree of the ML WSe$_2$-metal surfaces increases with the binding strength. The different bonding strength and band hybridization degree in different interfaces can be well explained by the $d$-band model.[34] Al has no $d$-orbitals and Ag and Au have full $d$-shells and they all bond with ML WSe$_2$ weakly, whereas Pt and Sc have open $d$-shells and bond with ML WSe$_2$ strongly. The relative position of $d$-band also plays an important role. As moving from the right to the left in the periodic table, the $d$-band moves up in energy. Although Pd and Ag both have full 4$d$-shells, the $d$-band of Pd is located nearer to $E_f$ than that of Ag thus strongly hybridizes with the valence bands of WSe$_2$. The $d$-band of Sc is located near the conduction bands of ML WSe$_2$ in energy. Therefore, the conduction bands of ML WSe$_2$ are perturbed more seriously than its valence bands when contacted with Sc.

To further study the contact natures in the vertical direction, the partial density of states (PDOS) of W and Se atoms in ML WSe$_2$-metal contacts are provided in Figure 3. After contacted with metal faces, there appear electronic states in the original band gap of ML WSe$_2$. PDOS at $E_f$ ($N(E_f)$) increases in this order: Au (0.39) < Al (0.50) < Ag (0.58) < Pt (1.08) < Pd (1.24) < Sc (2.1), a result consistent with the hybridization degree (for Sc contact, we compare its $3N(E_f)$ value with the $N(E_f)$ of other metal contacts because its interface unit cell area is 1/3 of others). The prominent overlap between Sc and WSe$_2$ in the original band gap of WSe$_2$ indicates a covalent bond formation between them, thus further confirming the absence of vertical Schottky barrier. In contrast, there are much fewer overlap states distributed in the original band gap for other contacts compared with those in the Sc contact. Because the overlap states near $E_f$ contribute to the electron or hole injections from the metal, the PDOS value at $E_f$ to a degree can reflect the quality of vertical contacts.

The electronic structures of free-standing BL WSe$_2$ and BL WSe$_2$-metal interfaces are shown in Figure 4, with a smaller indirect band gap of 1.43 eV for free-standing BL WSe$_2$. The band hybridization degree is similar to ML and can also be divided into the same three categories. The vertical Schottky barrier in BL WSe$_2$-Sc system is also absent due to the strong hybridization between the bands of BL WSe$_2$ and Sc. In BL WSe$_2$-Al, -Ag, -Pd and -Pt contacts, the type of the vertical Schottky barrier is same as the ML case; the Schottky barriers are $n$-type with $\Phi_V^e = 0.27$ and 0.20 eV in BL WSe$_2$-Al and -Ag, and are $p$-type with $\Phi_V^h = 0.27$ and 0.28 eV in BL WSe$_2$-Pd and -Pt, respectively. In these four metal contacts, $\Phi_V^e$ is decreased significantly in the $n$-type contacts while $\Phi_V^h$ is not altered much in the $p$-type contacts compared with those in the cases of ML. The vertical Schottky barrier changes from weak $n$-type in ML WSe$_2$-Au contact to weak $p$-type with $\Phi_V^h = 0.45$ eV in BL case. In both cases, $E_f$ is close to the band gap center of WSe$_2$. Experimentally, ambipolarity is observed in few layer WSe$_2$ FET with Au electrode, a result consistent with our calculation.[35]

WSe$_2$ hosts heavy 5$d$ elements with strong atomic SOC, much stronger than that in the more intensively studied MoS$_2$ system.[11] After inclusion of the SOC effects, the band structures of ML and BL WSe$_2$ are greatly modified as shown in the first and second panels of Figure 2 and 4 by lining up the bands with the vacuum level. The band gap of ML (BL) WSe$_2$ is reduced from 1.60(1.43) to 1.33(1.15) eV. The CBM of ML (BL) WSe$_2$ is changed slightly and only falls by 0.04 (-0.09) eV, but the VBM is significantly elevated by 0.23 (0.37) eV, after the SOC effects are included. Therefore, if the Fermi level pinning is absent, the electron vertical SBH is little affected, but the hole vertical SBH is decreased remarkably by about 0.23 (0.37) eV for ML (BL) case once the SOC effects are included. As shown in Figure 5, the CBMs of ML and BL WSe$_2$ is little changed (within 0.04 eV) in the Pd and Pt contacts when the SOC is included. By contrast, the VBM of ML (BL) WSe$_2$ is lifted by 0.04 (0.21) eV in ML (BL) WSe$_2$-Pd contact after inclusion of the SOC effect, and thus we have a reduced $\Phi_V^h$ = 0.18 (0.06) eV for hole injection. The VBM in ML (BL) WSe$_2$-Pd contact is not elevated as high as in free-standing ML (BL) WSe$_2$, reflecting a partial Fermi level pining during the SOC process. The reduction in $\Phi_V^h$ is especially remarkable in Pt contact. The VBM of WSe$_2$ touches $E_f$ in ML and BL WSe$_2$-Pt contacts after inclusion of the SOC effects because of the significant rise of the VBM by about 0.30 and 0.28 eV, respectively, leading to Ohmic or quasi-Ohmic interfaces. The rise amplitudes of the VBM in ML and BL WSe$_2$-Pt contacts are comparable with those in free-standing ML and BL WSe$_2$, implying a much depressed or even vanishing Fermi level pining during the SOC process in Pt contact. The dramatic change in the hole SBH indicates the importance of the SOC in determining the interfacal properties of 2D WSe$_2$-metal contacts and it is not ignorable. The predicted *p*-type quasi-Ohmic contact for ML and BL WSe$_2$-Pt appears in agreement with the recent experimental results of Javey *et al.* that the *p*-type WSe$_2$ FET in contact with metal Pt/Au/Pd electrode has almost linear output characteristics in the low-voltage regime.[6]

**Lateral SBH**

Because of the covalent bonding between WSe$_2$ and Sc, ML (BL) WSe$_2$-Sc can be regarded as a new metallic material. A lateral Schottky barrier is formed at the interface D, and its height $\Phi_L^{e/h}$ is determined by the energy difference between $E_f$ of the WSe$_2$-Sc complex system and the CBM (electron SBH) or VBM (hole SBH) of channel WSe$_2$. Because $\Phi_L^{e/h}$ depends on the absolute band edge positions of the semiconductor channel, the band edge positions of the semiconductor channel must be accurately determined. As has been testified in MoS$_2$-metal contacts, many-electron effect is greatly depressed by the charge transfer between channel TMD and the electrode and the doping caused by gate,[21] which significantly screen the electron-electron Coulomb interaction. As a rule of thumb, the transport gap of 2D TMD channel could be determined by the DFT band gap rather than the quasi-particle band gap. As shown in Figure S1, lateral *n*-type Schottky barriers are formed for ML and BL WSe$_2$-Sc contacts with electron SBHs $\Phi_L^{e(\text{Non-SOC})}$ = 0.29 and 0.16 eV and $\Phi_L^{e(\text{SOC})}$ = 0.25 and 0.25 eV, respectively.

**Fermi level pinning**

The partial Fermi level pinning is a synergic result of the metal work function modification and the interface states formation in the studied interface systems.[23] Fermi level pinning makes the contact nature complex and difficult to predict. We define the Fermi energy shift $\Delta E_f$ as

$$\Delta E_f = \begin{cases} E_D - E_f, \text{ for vertical Schottky barrier} \\ W - W_{WSe_2}, \text{ for lateral Schottky barrier} \end{cases} \quad (2)$$

where $E_D$ is the middle energy of the band gap of the WSe$_2$ adsorbed on metal substrates, $E_f$ is the Fermi level of the interfacial system, $W$ and $W_{WSe_2}$ are the work functions of the interfacial system and pristine ML or BL WSe$_2$, respectively. Negative (positive) $\Delta E_f$ means *n*-type (*p*-type) doping of WSe$_2$. $\Delta E_f$ as a function of the difference between the clean metal and ML (BL) WSe$_2$ work functions $W_M - W_{WSe_2}$ is shown in Figure S2. We applying a linear fit to all the data obtained with or without the SOC effects. The slope is 0.40 in both ML and BL WSe$_2$-metal contacts. The slope close to 1 means no Fermi level pinning and close to 0 indicates a strong Fermi level pinning. We therefore observe a partial Fermi level pinning picture in the six ML (BL) WSe$_2$-metals contacts.

**Tunneling barrier**

The tunneling barrier is another figure of merit to evaluate a contact, here its height $\Delta V$ defined as the potential energy above $E_f$ between the WSe$_2$ and metal surfaces, indicated by the black rectangular in Figure 6, and its width $w_B$ defined as the full width at half maximum of the $\Delta V$. As shown in Figure 6 and Table 2, the tunneling barrier is insensitive to the number of WSe$_2$ layers. The weak bonding (Al, Ag and Au) and medium bonding (Pd and Pt) interfaces have a notably high $\Delta V$ and a notably wide $w_B$. On the contrary, there is no tunneling barrier at the strong bonding interfaces (Sc, and Ti), indicating a higher electron injection efficiency and thus a lower contact resistance. We estimate the tunneling probabilities $T_B$ from metal to WSe$_2$ using a square potential barrier model as:

$$T_B = \exp(-2 \times \frac{\sqrt{2m\Delta V}}{h} \times \omega_B) \quad (3)$$

where $m$ is the effective mass of a free electron and $\hbar$ is the Plank's constant. The $T_B$ values are thus estimated to be 100 (100), 26.9 (47.9), 39.9 (52.2), 34.3 (42.7), 52.4 (78.8) and 31.0 (41.0)% for ML (BL) WSe$_2$-Sc, Al, Ag, Au, Pd and Pt contacts, respectively. Apparently, Sc contacts have a perfect tunneling transmission.

**Quantum transport simulation**

A more direct and reliable approach to determine the SBH of a 2D WSe$_2$ transistor is the *ab initio* quantum transport simulation. As an example, in Figure 7(a), we present the simulated transport properties of ML WSe$_2$ transistor with Pt electrodes (the SOC is not included). The length of the channel is 73 Å. From the transmission spectrum shown in Figure 7(b), we can see a transport gap of 1.65 eV and a hole SBH of 0.34 eV. From the local density of states plotted in Figure 7(c), a band gap of ~1.8 eV in the free ML WSe$_2$ region (dark blue) and a hole SBH of ~ 0.35 eV are clearly visible, comparable with the corresponding values derived from the transmission spectrum. The band bending of the free ML WSe$_2$ near the interface is not sharp, showing a weak built-in electric field between the source/drain and channel region. The SBH calculated in the transport simulation (0.34-0.35 eV) is also close to that (0.29 eV) calculated from the energy band analysis without inclusion of the SOC.

This good agreement indicates that the energy band calculation is suitable in describing the vertical SBH in the weak bonding contacts although it often gives an artificial vanishing lateral SBH in the strong bonding contacts.[21, 32] The reason lies in that in the vertical SBH calculation of a weak bonding contact, the coupling between the metal surface and the semiconductor is taken into account where the two parts are treated as a whole, but in the lateral SBH calculation of a strong bonding contact, the coupling between the source/drain region (metal and metal contacted 2D semiconductor) and the channel semiconductor is not considered where they are treated separately. We therefore believe that the Ohmic or quasi-Ohmic feature of ML/BL WSe$_2$-Pt contacts should be kept in a quantum transport simulation with inclusion of the SOC though it is unavailable now.

To provide a clear picture, the SBHs obtained by different methods are summarized in Figure 8. When contacted with the same metal, compared with those in ML case, the electron and hole SBHs in BL case both tend to be decreased due to the much reduced band gap. From left to right, the ML and BL WSe$_2$ are gradually changed from *n*- to *p*-type doping, which can be utilized to build *p*−*n* junctions, the most fundamental device building blocks for diverse optoelectronic functions. A ML WSe$_2$ device with Ti as cathode and Pd as anode is synthesized and can serve as a solar cell, photodiode, and light-emitting diode with impressive performances.[9] *P*-type ML WSe$_2$-Pd contact has a smaller SBH compared with Ag and Au contacts, which is well consistent with the contact resistance measurement.[5] In the absence of the SOC, Pd contact has the smallest hole SBH with a value no less than 0.22 eV though Pt has a larger work function than Pd (5.76 eV vs 5.36 eV in our work as given in Table 1). However, Pt contact wins the smallest hole SBH (actually 0 eV) in the presence of the SOC.

In the light of Schottky barrier and tunneling barrier, the nature of investigated ML WSe$_2$- metal contacts can be classified into four types as summarized in Figure 9. ML WSe$_2$-Sc contact is type I with a vanishing tunneling barrier and a finite *n*-type lateral Schottky barrier (Figure 9a). A non-zero tunneling barrier exists in the rest three types of contacts. Schottky barrier is formed at the interface B in type II (*n*-type) and III (*p*-type). The nature of ML WSe$_2$-Al, -Ag and -Au belongs to type II (Figure 9b) and that of ML WSe$_2$-Pd belongs to type III (Figure 9c). Type IV (ML WSe$_2$-Pt) can be expected as an excellent contact interface with an Ohmic or quasi-Ohmic contact (Figure 9d). However, the tunneling barrier with moderate tunneling probabilities (31.0% for Pt) would degrade its performance. As for BL WSe$_2$, BL WSe$_2$-Sc, Al, Ag, Pd and Pt interfaces keep the same contact type as the corresponding ML ones. However, WSe$_2$-Au changes into type IV contact in BL case.

**Conclusion**

We provide the first comparative study of the interfacial properties of ML and BL WSe$_2$ on Sc, Al, Ag, Au, Pd, and Pt surfaces by using *ab initio* energy band calculations with inclusion of the SOC effects and a dual interface model. Compared with ML WSe$_2$-metal contacts, the electron and hole SBHs are decreased in BL WSe$_2$-metal contacts due to the smaller band gap in BL WSe$_2$, and the polarity of WSe$_2$-Au contact changes from *n*-type to *p*-type. The hole SBH is greatly reduced by the SOC effects in both ML and BL WSe$_2$-metal contacts. In the absence of the SOC, Pd contact has the smallest hole SBH with a value no less than 0.22 eV. Dramatically, *p*-type Ohmic or quasi-Ohmic contact appears in ML and BL WSe$_2$-Pt interfaces with inclusion of the SOC. *Ab initio* quantum transport simulation gives a similar SBH for ML WSe$_2$-Pt interface in the absence of the SOC. This fundamental study gives a deep insight into 2D WSe$_2$-metal interfaces and provides a theoretical foundation for the selection of metal electrodes in WSe$_2$ devices.

**Methods**

The geometry optimizations are carried out by employing the CASTEP package[36] with the ultrasoft pseudopotential[37] and plane-wave basis set. The cut-off energy is 400 eV. To take the dispersion interaction into account, a DFT-D semiempirical dispersion-correction approach is used with the Ortmann-Bechstedt-Schmidt (OBS) scheme.[38] The dipole correction to the total energies is adopted. The stopping criteria for the ionic relaxation are such that the remnant force on each atom is below 0.01 eV/Å and that energies are converged to within $10^{-5}$ eV per atom. The electronic structures are calculated with the projector-augmented wave (PAW) pseudopotential[39, 40] and plane-wave basis set with a cut-off energy of 400 eV implemented in the Vienna *ab initio* simulation package (VASP) in order to analyze the band components.[41-44] The Monkhorst-Pack[45] *k*-point mesh is sampled with a separation of about 0.10 and 0.03 Å$^{-1}$ in the Brillouin zone during the relaxation and electronic calculation periods, respectively. Our tests show that the band structures generated from CASTEP and VASP packages coincide well.

The WSe$_2$ transistor is simulated using the DFT coupled with the nonequilibrium Green's function (NEGF) method, as implemented in the ATK 11.8 package [46, 47]. The single-zeta plus polarization (SZP) basis set is used. The Monkhorst-Pack *k*-point meshes for the central region and electrodes are sampled with 1×50×1 and 50×50×1 separately. The temperature is set to 300 K. The Neumann condition is used on the boundaries of the direction vertical to the WSe$_2$ plane. On the surfaces connecting the electrodes and the central region, we employ Dirichlet boundary condition to ensure the charge neutrality in the source and the drain region. The transmission coefficient at energy *E* averaged over 50 $k_y$-points perpendicular to the transport direction (*x* direction) is obtain by

$$T(E) = \mathrm{Tr}[G^r \Gamma_L(E) G^a \Gamma_R(E)] \quad (4)$$

Where $G^{r(a)}$ is the retarded (advanced) Green function and $\Gamma_{L(R)}(E) = i(\Sigma^r_{L(R)} - \Sigma^a_{L(R)})$ is the level broadening due to left (right) electrode expressed in terms of the electrode self-energy $\Sigma_{L(R)}$. Throughout the paper, the generalized gradient approximation (GGA) functional to the exchange-correction functional of the Perdew–Wang 91 (PW91) form[46] is adopted.

**Acknowledgements**

This work was supported by the National Natural Science Foundation of China (Nos. 11274016, 11474012, 11047018, and 60890193), the National Basic Research Program of China (Nos. 2013CB932604 and 2012CB619304), Fundamental Research Funds for the Central Universities, and National Foundation for Fostering Talents of Basic Science (No. J1030310/No. J1103205).

**Additional information**

**Competing financial interests:** The authors declare no competing financial interest.

**Corresponding Author**

Correspondence to: Jing Lu (jinglu@pku.edu.cn)

# References

1. Wang, Q. H.; Kalantar-Zadeh, K.; Kis, A.; Coleman, J. N.; Strano, M. S. Electronics and Optoelectronics of Two-dimensional Transition Metal Dichalcogenides. *Nature Nanotech.* **2012,** *7* (11), 699-712.
2. Radisavljevic, B.; Radenovic, A.; Brivio, J.; Giacometti, V.; Kis, A. Single-layer $MoS_2$ Transistors *Nature Nanotech.* **2011,** *6* (3), 147-150.
3. Das, S.; Chen, H.-Y.; Penumatcha, A. V.; Appenzeller, J. High Performance Multilayer $MoS_2$ Transistors with Scandium Contacts. *Nano Lett.* **2013,** *13* (1), 100-105.
4. Liu, W.; Kang, J.; Sarkar, D.; Khatami, Y.; Jena, D.; Banerjee, K. Role of Metal Contacts in Designing High-Performance Monolayer n-Type $WSe_2$ Field Effect Transistors. *Nano Lett.* **2013,** *13* (5), 1983-1990.
5. Fang, H.; Chuang, S.; Chang, T. C.; Takei, K.; Takahashi, T.; Javey, A. High-Performance Single Layered $WSe_2$ p-FETs with Chemically Doped Contacts. *Nano Lett.* **2012,** *12* (7), 3788-3792.
6. Tosun, M.; Chuang, S.; Fang, H.; Sachid, A. B.; Hettick, M.; Lin, Y.; Zeng, Y.; Javey, A. High-Gain Inverters Based on $WSe_2$ Complementary Field-Effect Transistors. *ACS Nano* **2014,** *8* (5), 4948-4953.
7. Zhao, W.; Ribeiro, R. M.; Eda, G. Electronic Structure and Optical Signatures of Semiconducting Transition Metal Dichalcogenide Nanosheets. *Acc. Chem. Res.* **2014,** *48* (1), 91-99.
8. Ross, J. S.; Klement, P.; Jones, A. M.; Ghimire, N. J.; Yan, J.; Mandrus, D. G.; Taniguchi, T.; Watanabe, K.; Kitamura, K.; Yao, W.; Cobden, D. H.; Xu, X. Electrically Tunable Excitonic Light-emitting Diodes Based on Monolayer $WSe_2$ p-n Junctions. *Nature Nanotech.* **2014,** *9* (4), 268-272.
9. Pospischil, A.; Furchi, M. M.; Mueller, T. Solar-energy Conversion and Light Emission in an Atomic Monolayer p-n Diode. *Nature Nanotech.* **2014,** *9* (4), 257-261.
10. Neal, A. T.; Liu, H.; Gu, J.; Ye, P. D. Magneto-transport in $MoS_2$: Phase Coherence, Spin–Orbit Scattering, and the Hall Factor. *ACS Nano* **2013,** *7* (8), 7077-7082.
11. Xiao, D.; Liu, G.-B.; Feng, W.; Xu, X.; Yao, W. Coupled Spin and Valley Physics in Monolayers of $MoS_2$ and Other Group-VI Dichalcogenides. *Phys. Rev. Lett.* **2012,** *108* (19), 196802.
12. Yuan, H.; Bahramy, M. S.; Morimoto, K.; Wu, S.; Nomura, K.; Yang, B.-J.; Shimotani, H.; Suzuki, R.; Toh, M.; Kloc, C.; Xu, X.; Arita, R.; Nagaosa, N.; Iwasa, Y. Zeeman-type Spin Splitting Controlled by an Electric Field. *Nature Phys.* **2013,** *9* (9), 563-569.
13. Xu, X.; Yao, W.; Xiao, D.; Heinz, T. F. Spin and Pseudospins in Layered Transition Metal Dichalcogenides. *Nature Phys.* **2014,** *10* (5), 343-350.
14. Jones, A. M.; Yu, H.; Ghimire, N. J.; Wu, S.; Aivazian, G.; Ross, J. S.; Zhao, B.; Yan, J.; Mandrus, D. G.; Xiao, D.; Yao, W.; Xu, X. Optical Generation of Excitonic Valley Coherence in Monolayer $WSe_2$. *Nature Nanotech.* **2013,** *8* (9), 634-638.
15. Mak, K. F.; McGill, K. L.; Park, J.; McEuen, P. L. The Valley Hall Effect in $MoS_2$ Transistors. *Science* **2014,** *344* (6191), 1489-1492.
16. Lee, C.-H.; Lee, G.-H.; van der Zande, A. M.; Chen, W.; Li, Y.; Han, M.; Cui, X.; Arefe, G.; Nuckolls, C.; Heinz, T. F.; Guo, J.; Hone, J.; Kim, P. Atomically Thin p–n Junctions with Van der Waals Heterointerfaces. *Nature Nanotech.* **2014,** *9* (9), 676-681.
17. Sie, E. J.; McIver, J. W.; Lee, Y.-H.; Fu, L.; Kong, J.; Gedik, N. Valley-selective Optical Stark Effect in Monolayer $WS_2$. *Nature Mater.* **2015,** *14*, 290–294.
18. Kim, J.; Hong, X.; Jin, C.; Shi, S.-F.; Chang, C.-Y. S.; Chiu, M.-H.; Li, L.-J.; Wang, F. Ultrafast Generation of Pseudo-magnetic Field for Valley Excitons in $WSe_2$ Monolayers. *Science* **2014,** *346* (6214), 1205-1208.
19. Kang, J.; Liu, W.; Sarkar, D.; Jena, D.; Banerjee, K. Computational Study of Metal Contacts to Monolayer Transition-Metal Dichalcogenide Semiconductors. *Phys. Rev. X* **2014,** *4* (3), 031005.
20. Cheng, R.; Li, D.; Zhou, H.; Wang, C.; Yin, A.; Jiang, S.; Liu, Y.; Chen, Y.; Huang, Y.; Duan, X. Electroluminescence and Photocurrent Generation from Atomically Sharp $WSe_2/MoS_2$ Heterojunction p–n Diodes. *Nano Lett.* **2014,** *14* (10), 5590-5597.
21. Zhong, H.; Ni, Z.; Wang, Y.; Ye, M.; Song, Z.; Pan, Y.; Quhe, R.; Yang, J.; Yang, L.; Shi, J.; Lu, J. Interfacial Properties of Monolayer and Bilayer $MoS_2$ Contacts with Metals: Depressed Many-electron Effects. *arXiv:1501.01071 [cond-mat.mes-hall]* **2015**.
22. Pan, Y.; Wang, Y.; Wang, L.; Zhong, H.; Quhe, R.; Ni, Z.; Ye, M.; Mei, W.-N.; Shi, J.; Guo, W.;

Yang, J.; Lu, J. Graphdiyne-metal Contacts and Graphdiyne Transistors. *Nanoscale* **2015,** *7* (5), 2116-2127.
23. Gong, C.; Colombo, L.; Wallace, R. M.; Cho, K. The Unusual Mechanism of Partial Fermi Level Pinning at Metal–$MoS_2$ Interfaces. *Nano Lett.* **2014,** *14* (4), 1714-1720.
24. Kang, J.; Liu, W.; Banerjee, K. High-performance $MoS_2$ Transistors with Low-resistance Molybdenum Contacts. *Appl. Phys. Lett.* **2014,** *104* (9), 093106.
25. Chen, W.; Santos, E. J. G.; Zhu, W.; Kaxiras, E.; Zhang, Z. Tuning the Electronic and Chemical Properties of Monolayer $MoS_2$ Adsorbed on Transition Metal Substrates. *Nano Lett.* **2013,** *13* (2), 509-514.
26. Popov, I.; Seifert, G.; Tománek, D. Designing Electrical Contacts to $MoS_2$ Monolayers: A Computational Study. *Phys. Rev. Lett.* **2012,** *108* (15), 156802.
27. Wei, L.; Jiahao, K.; Wei, C.; Sarkar, D.; Khatami, Y.; Jena, D.; Banerjee, K. High-performance Few-layer-$MoS_2$ Field-effect-transistor with Record Low Contact-resistance, IEEE International Electron Devices Meeting (IEDM), Washington DC, Dec. 9-11, 2013, 499–502
28. Amani, M.; Chin, M. L.; Birdwell, A. G.; O'Regan, T. P.; Najmaei, S.; Liu, Z.; Ajayan, P. M.; Lou, J.; Dubey, M. Electrical Performance of Monolayer $MoS_2$ Field-effect Transistors Prepared by Chemical Vapor Deposition. *Appl. Phys. Lett.* **2013,** *102* (19), 193107.
29. Yun, W. S.; Han, S. W.; Hong, S. C.; Kim, I. G.; Lee, J. D. Thickness and Strain Effects on Electronic Structures of Transition Metal Dichalcogenides: 2H-$MX_2$ Semiconductors (M= Mo, W; X= S, Se, Te). *Phys. Rev. B* **2012,** *85* (3), 033305.
30. Al-Hilli, A. A.; Evans, B. L. The Preparation and Properties of Transition Metal Dichalcogenide Single Crystals. *J. Cryst. Growth* **1972,** *15* (2), 93-101.
31. Liu, Q.; Li, L.; Li, Y.; Gao, Z.; Chen, Z.; Lu, J. Tuning Electronic Structure of Bilayer $MoS_2$ by Vertical Electric Field: A First-Principles Investigation. *J. Phys. Chem. C* **2012,** *116* (40), 21556-21562.
32. Pan, Y.; Wang, Y.; Ye, M.; Quhe, R.; Zhong, H.; Song, Z.; Peng, X.; Li, J.; Yang, J.; Shi, J.; Lu, J. Monolayer Phosphorene-Metal Interfaces. **2015,** in preparation.
33. Qiu, H.; Pan, L.; Yao, Z.; Li, J.; Shi, Y.; Wang, X. Electrical Characterization of Back-gated Bi-layer $MoS_2$ Field-effect Transistors and the Effect of Ambient on Their Performances. *Appl. Phys. Lett.* **2012,** *100* (12), 123104.
34. Zheng, J.; Wang, Y.; Wang, L.; Quhe, R.; Ni, Z.; Mei, W.-N.; Gao, Z.; Yu, D.; Shi, J.; Lu, J. Interfacial Properties of Bilayer and Trilayer Graphene on Metal Substrates. *Sci. Rep.* **2013,** *3*, 2081, doi:10.1038/srep02081.
35. Fang, H.; Tosun, M.; Seol, G.; Chang, T. C.; Takei, K.; Guo, J.; Javey, A. Degenerate n-Doping of Few-Layer Transition Metal Dichalcogenides by Potassium. *Nano Lett.* **2013,** *13* (5), 1991-1995.
36. Clark, S. J.; Segall, M. D.; Pickard, C. J.; Hasnip, P. J.; Probert, M. J.; Refson, K.; Payne, M. C. First Principles Methods Using CASTEP. *Z. Kristallogr.* **2005,** *220* (5-6), 567-570.
37. Vanderbilt, D. Soft Self-consistent Pseudopotentials in a Generalized Eigenvalue Formalism. *Phys. Rev. B* **1990,** *41* (11), 7892-7895.
38. Ortmann, F.; Bechstedt, F.; Schmidt, W. G. Semiempirical Van der Waals Correction to the Density Functional Description of Solids and Molecular Structures. *Phys. Rev. B* **2006,** *73* (20), 205101.
39. Blochl, P. E. Projector Augmented-wave Method. *Phys. Rev. B* **1994,** *50* (24), 17953-17979.
40. Kresse, G.; Joubert, D. From Ultrasoft Pseudopotentials to the Projector Augmented-wave Method. *Phys. Rev. B* **1999,** *59* (3), 1758-1775.
41. Kresse, G.; Hafner, J. Ab Initio Molecular-dynamics For Liquid-metals. *Phys. Rev. B* **1993,** *47* (1), 558-561.
42. Kresse, G.; Hafner, J. Ab-initio Molecular-dynamics Simulation of the Liquid-metal Amorphous-semiconductor Transition in Germaniun. *Phys. Rev. B* **1994,** *49* (20), 14251-14269.
43. Kresse, G.; Furthmuller, J. Efficiency of Ab-initio Total Energy Calculations for Metals and Semiconductors Using a Plane-wave Basis Set. *Comput. Mater. Sci.* **1996,** *6* (1), 15-50.
44. Kresse, G.; Furthmuller, J. Efficient Iterative Schemes for Ab Initio Total-energy Calculations Using a Plane-wave Basis Set. *Phys. Rev. B* **1996,** *54* (16), 11169-11186.
45. Monkhorst, H. J.; Pack, J. D. Special Points for Brillouin-zone Integrations. *Phys. Rev. B* **1976,** *13* (12), 5188-5192.
46. Perdew, J. P.; Wang, Y. Accurate and Simple Analytic Representation of the Electron-gas

Correlation Energy. *Phys. Rev. B* **1992,** *45* (23), 13244-13249.

**Table 1.** Calculated interfacial properties of ML and BL WSe$_2$ on the metal surfaces. $a_{hex}^{exp}$ represents the experimental cell parameters of the surface unit cells shown in Figure 1 for various metals, with lattice mismatch in percentage given below in parenthesis. The equilibrium distance $d_z$ is the averaged distance between the surface Se atoms of WSe$_2$ and the relaxed positions of the topmost metal layer in the $z$ direction. $E_b$ is the binding energy per surface W atom between WSe$_2$ and a given metal. $W_M$ and $W$ are the calculated work functions for clean metal surface and WSe$_2$-metal contact, respectively. The SBHs obtained from band calculation with (without) inclusion of the SOC, transport simulation without inclusion of the SOC, and obtained in the previous work without inclusion of the SOC are given for comparison. Electron SBH is given for *n*-type Schottky barrier and hole SBH is given for *p*-type Schottky barrier. The Schottky barrier is always formed at the vertical direction except for Sc surface.

| Metal | $a_{hex}^{exp}$ (Å) | $W_M$ (eV) | ML WSe$_2$ | | | | BL WSe$_2$ | | | |
|---|---|---|---|---|---|---|---|---|---|---|
| | | | $d_z$ (Å) | $E_b$ (eV) | $W$ (eV) | SBH (eV) | $d_z$ (Å) | $E_b$ (eV) | $W$ (eV) | SBH (eV) |
| Sc | 3.308 (0.91%) | 3.60 | 2.736 | 0.918 | 3.75 | 0.29[n] (0.25[n])[S] | 2.512 | 1.049 | 3.94 | 0.16[n] (0.25[n])[S] |
| Al | 5.720 (0.68%) | 4.12 | 2.959 | 0.288 | 4.15 | 0.62[n] | 2.885 | 0.367 | 4.16 | 0.27[n] |
| Ag | 5.778 (1.74%) | 4.49 | 2.693 | 0.302 | 4.26 | 0.40[n] | 2.684 | 0.240 | 4.56 | 0.20[n] |
| Au | 5.768 (1.39%) | 5.23 | 2.712 | 0.182 | 4.71 | 0.53[n] (0.70[n])[a] | 2.773 | 0.160 | 4.85 | 0.45[p] |
| Pd | 5.500 (3.19%) | 5.36 | 2.395 | 0.602 | 4.84 | 0.22[p] (0.35[p])[a] (0.18[p])[S] | 2.271 | 0.706 | 5.05 | 0.27[p] (0.06[p])[S] |
| Pt | 5.549 (2.48%) | 5.76 | 2.652 | 0.525 | 5.22 | 0.30[p] (0.34[p])[T] (0.00[p])[S] | 2.770 | 0.597 | 5.21 | 0.28[p] (0.00[p])[S] |

[a] From Ref. [19].
[n] *n*-type Schottky barrier.
[p] *p*-type Schottky barrier.
[S] In the presence of the SOC.
[T] Value from the transport simulation.

**Table 2.** Tunneling barrier height $\Delta V$, width $w_B$, and probabilities ($T_B$) through the ML/BL WSe$_2$.

| Metal | ML WSe$_2$ | | | BL WSe$_2$ | | |
|---|---|---|---|---|---|---|
| | $\Delta V$ (eV) | $w_B$ (Å) | $T_B$ (%) | $\Delta V$ (eV) | $w_B$ (Å) | $T_B$ (%) |
| Sc | 0.000 | 0.000 | 100 | 0.000 | 0.000 | 100 |
| Al | 2.284 | 0.850 | 26.9 | 0.996 | 0.720 | 47.9 |
| Ag | 1.787 | 0.670 | 39.9 | 0.87 | 0.68 | 52.2 |
| Au | 2.199 | 0.704 | 34.3 | 1.282 | 0.734 | 42.7 |
| Pd | 1.477 | 0.518 | 52.4 | 0.555 | 0.313 | 78.8 |
| Pt | 2.533 | 0.716 | 31.0 | 1.420 | 0.731 | 41.0 |

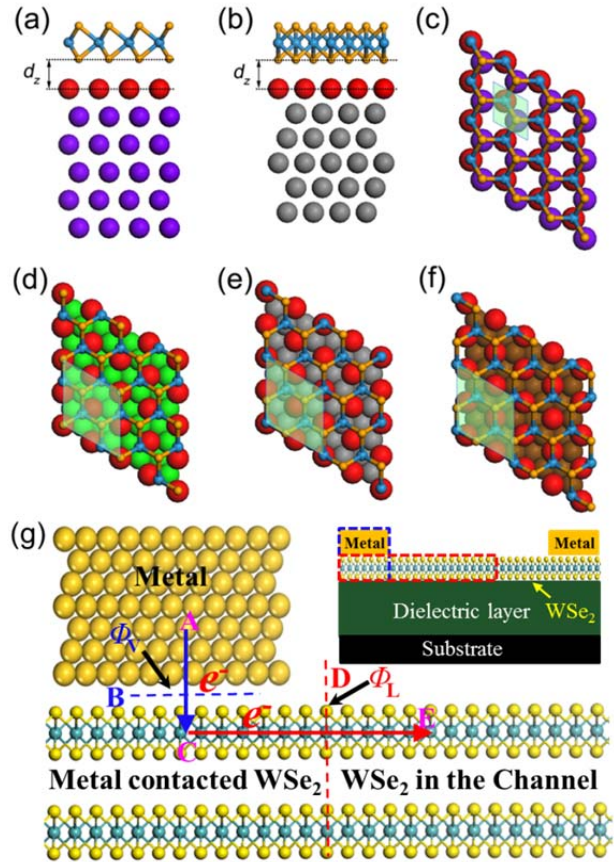

**Figure 1.** Interfacial structures of the most stable configuration for ML WSe$_2$ on metal surfaces. Side views of (a) WSe$_2$ on Sc(0001) surface and (b) on other metal surfaces. Top views of contacts (c) Sc-WSe$_2$, (d) Al/Pt-WSe$_2$, (e) Pd-WSe$_2$, (f) Ag/Au-WSe$_2$. $d_z$ is the equilibrium distance between the metal surface and the bottom layer WSe$_2$. The rhombi plotted in light green shadow shows the unit cell for each structure. (g) Schematic cross-sectional view of a typical metal contact to intrinsic WSe$_2$. A, C, and E denote the three regions while B and D are the two interfaces separating them. Blue and red arrows show the pathway (A→B→C→D→E) of electron injection from contact metal (A) to the WSe$_2$ channel (E). Inset figure shows the typical topology of a WSe$_2$ field effect transistor.

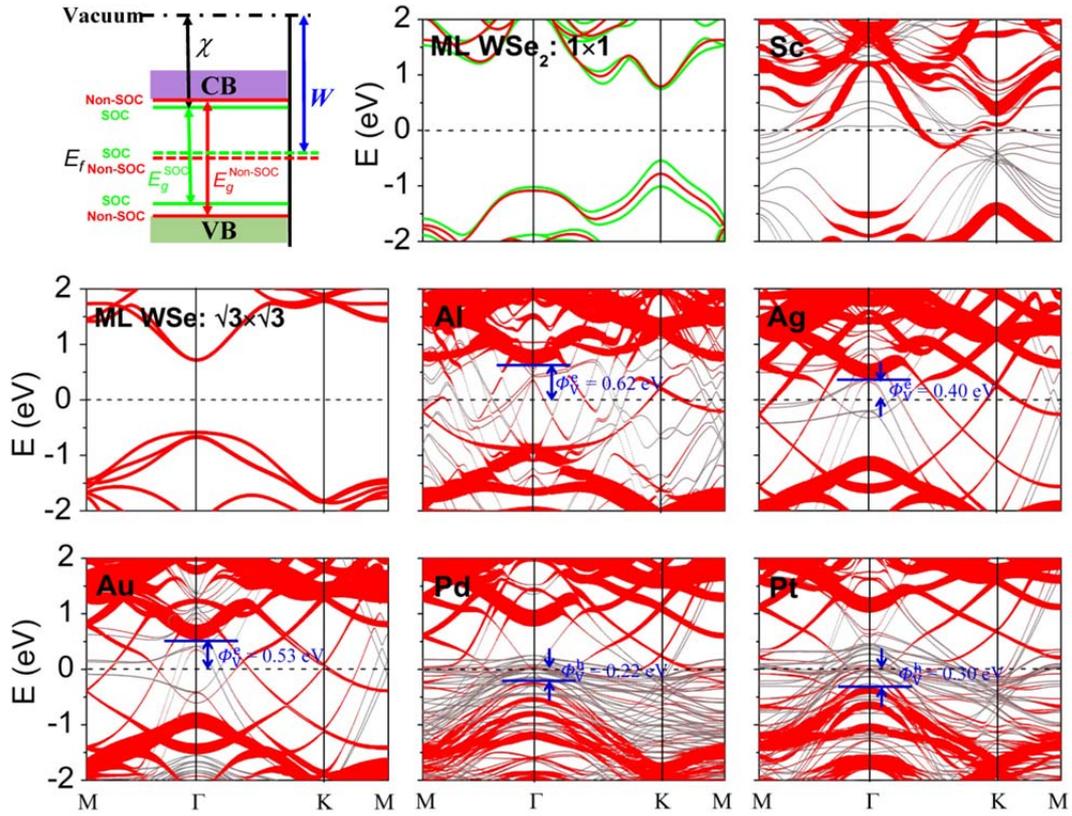

**Figure 2.** First panel: Schematic illustration of the absolute band positions with respect to the vacuum level by the DFT method with and without inclusion of the SOC effects for ML WSe$_2$. The rest: Band structures of ML WSe$_2$ and ML WSe$_2$-Sc, -Al, -Ag, -Au, -Pt, and -Pt contacts, respectively. Gray line: metal surface bands; red line: bands of WSe$_2$ without considering the SOC effects. The line width is proportional to the weight. Green line: bands of WSe$_2$ with the SOC effects. The Fermi level is set at zero.

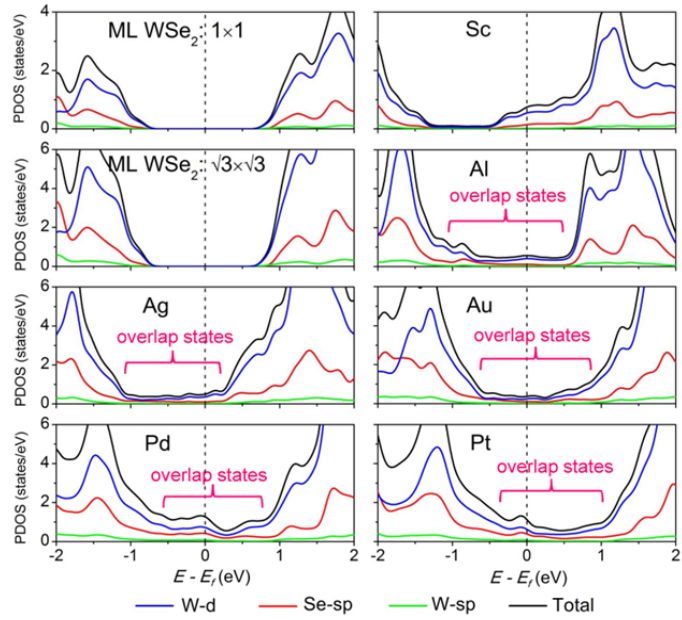

**Figure 3.** Partial density of states (PDOS) of W and Se electron orbitals, for ML WSe$_2$, ML WSe$_2$-Sc, -Al, -Ag, -Au, -Pd, and -Pt systems, respectively, in the absence of the SOC. The blue, red, green, and black curves represent *d*-orbital of W atoms, *sp*-orbital of Se atoms, *sp*-orbital of W atoms, and the total PDOS as indicated by the legend below the plot. The Fermi level is set at zero.

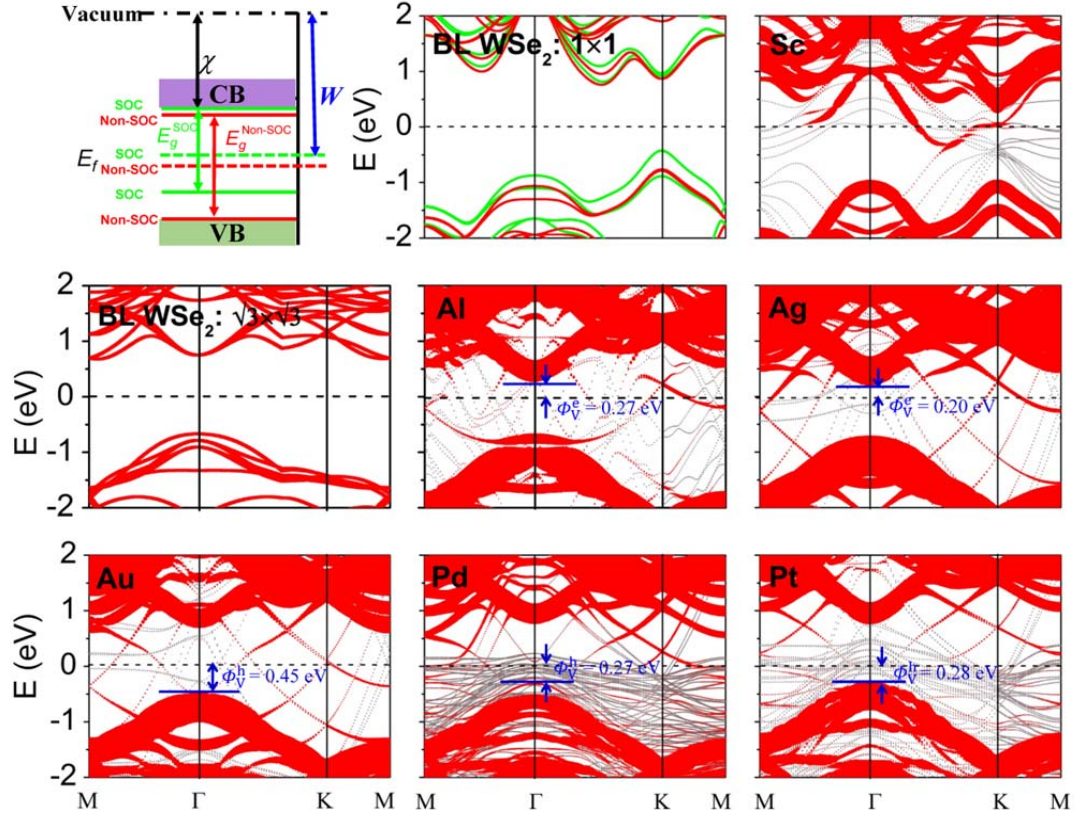

**Figure 4.** First panel: Schematic illustration of the absolute band positions with respect to the vacuum level by the DFT method with and without inclusion of SOC effects for BL WSe$_2$. The rest: Band structures of BL WSe$_2$ and BL WSe$_2$-Sc, -Al, -Ag, -Au, -Pt, and -Pt contacts, respectively. Gray line: metal surface bands; red line: bands of WSe$_2$ without considering the SOC effects. The line width is proportional to the weight. Green line: bands of WSe$_2$ with the SOC effects. The Fermi level is set at zero.

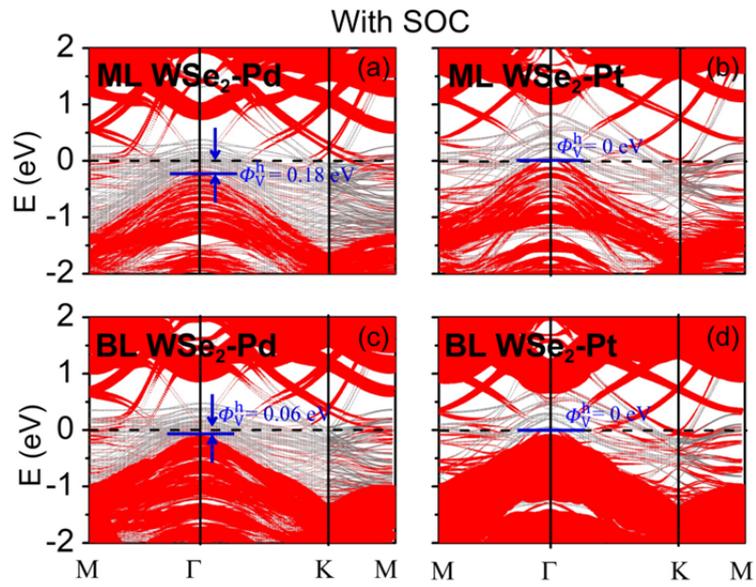

**Figure 5.** Band structures of (a,b) ML and (c,d) BL WSe$_2$ on (a,c) Pd, and (b,d) Pt surfaces with the SOC effects, respectively. Gray line: metal surface bands; red line: bands of WSe$_2$. The line width is proportional to the weight. The Fermi level is set at zero.

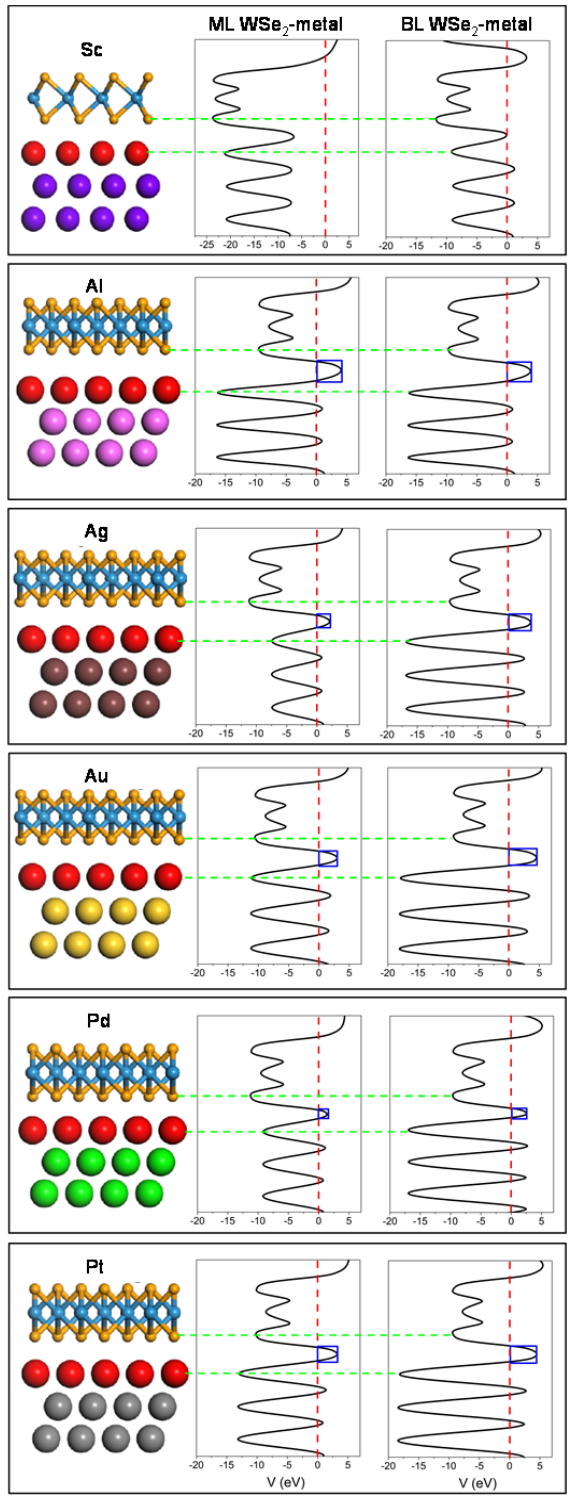

**Figure 6.** Average electrostatic potential $V$ in planes normal to the ML and BL WSe$_2$-metal contacts. The red dash lines represent the Fermi level.

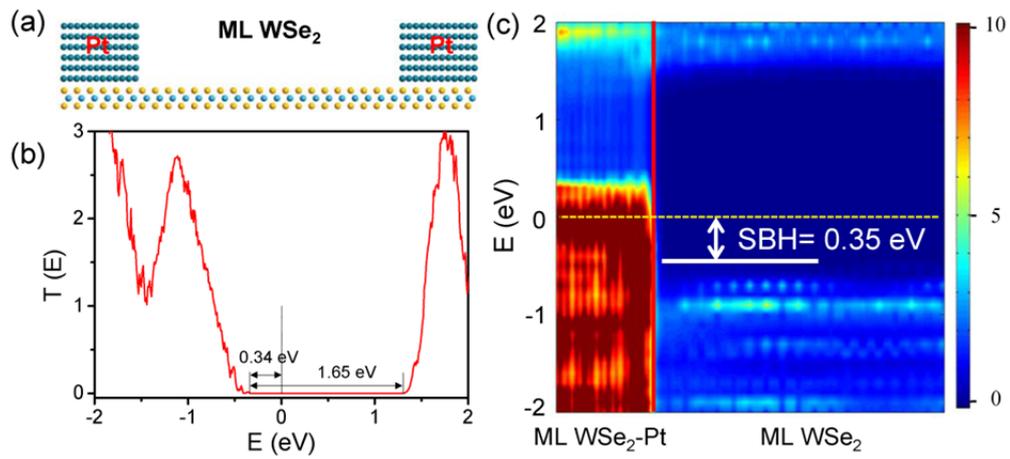

**Figure 7.** Simulation of a ML WSe$_2$ transistor with Pt as electrodes without inclusion of the SOC. (a) Schematic configuration. (b) Zero-bias transmission spectrum. The transport gap and hole SBH are indicated. (c) Local density of states (LDOS) in color coding for the device. The red line indicates the boundary of ML WSe$_2$/Pt and the free-standing ML WSe$_2$, and the yellow dashed line indicates the Fermi level.

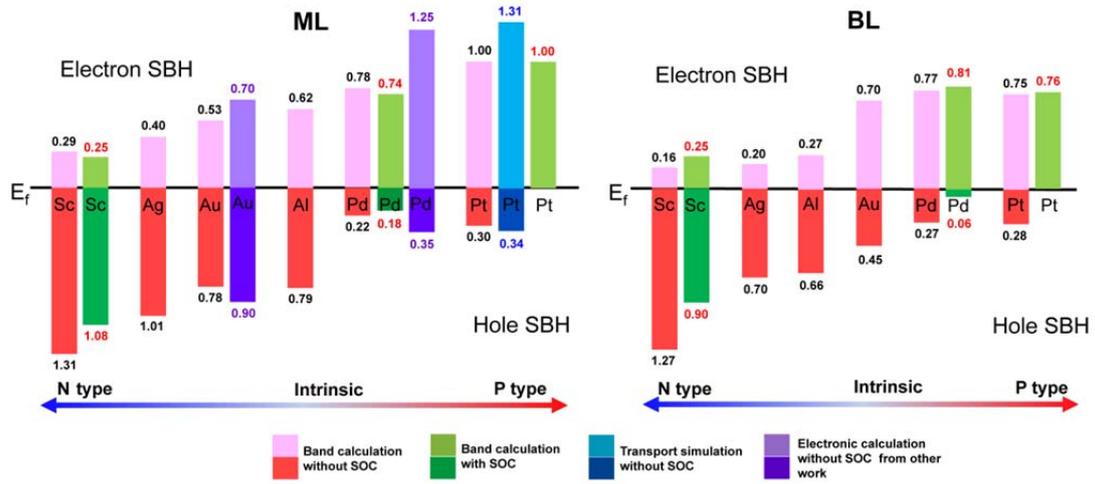

**Figure 8.** Electron and hole SBHs of (a) ML and (b) BL WSe$_2$-Sc, Al, Ag, Au, Pd and Pt contacts. The light (deep) red, green, blue, and purple rectangle present the electron (hole) SBH obtained from band calculation without the SOC, band calculation with the SOC, transport simulation without the SOC, and data from Ref. [19], respectively.

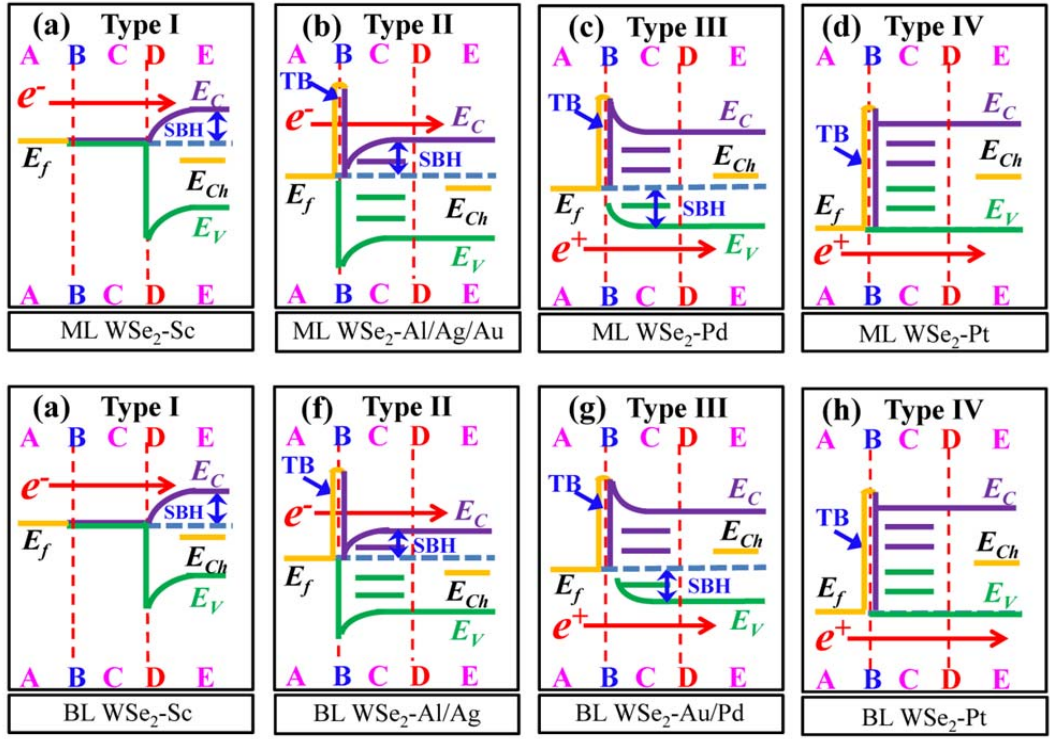

**Figure 9.** (a)-(h) Eight band diagrams of Figure 1(g), depending on the type of metals and WSe$_2$ layer number. Examples are provided at the bottom of each diagram. $E_f$ and $E_{Ch}$ denote the Fermi level of the absorbed system and intrinsic channel WSe$_2$, respectively. Red arrows indicate the direction of electron or hole flow.

TOC:

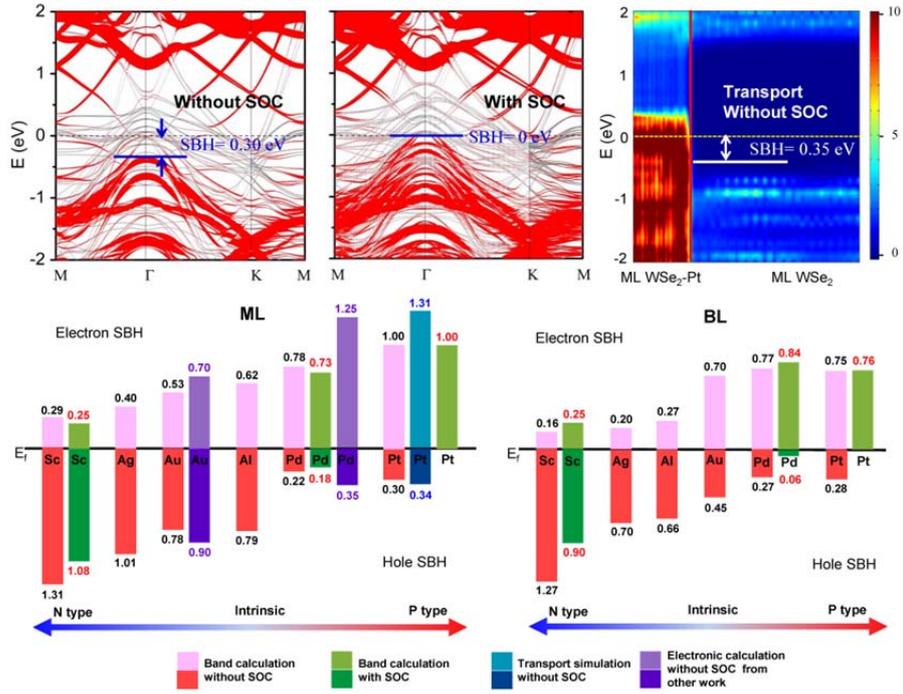